# Local structure surrounding V sites in Co doped $ZnV_2O_4$

P. Shahi[1], N. Tiwari[2], D. Bhattacharyya[2], S. N. Jha[2], A. K. Ghosh[3], A. Banerjee[4], H. Singh[5], A.K. Sinha[5] and Sandip Chatterjee[1,*]

[1)] Department of Physics, Indian Institute of Technology (Banaras Hindu University), Varanasi-221 005, India
[2)] Atomic and Molecular Physics Division, Bhabha Atomic Research Centre, Mumbai – 400 085, India
[3)] Department of Physics, Banaras Hindu University, Varanasi-221 005, India
[4)] UGC-DAE Consortium for Scientific Research, Indore, Madhya Pradesh 452017, India
[5)] Synchrotron Utilization Division Centre for Advanced Technology, Indore 452013, India

**Abstract**

Co doped $ZnV_2O_4$ has been investigated by Synchrotron X-ray diffraction, Magnetization measurement and Extended X-ray absorption fine structure (EXAFS) analysis. With Co doping in the Zn site the system moves towards the itinerant electron limit. From Synchrotron and magnetization measurement it is observed that there is an effect in bond lengths and lattice parameters around the magnetic transition temperature. The EXAFS study indicates that Co ion exists in the High spin state in Co doped $ZnV_2O_4$.

## INTRODUCTION:

$ZnV_2O_4$ is one of the Mott-insulators which lies at the boundary of Mott-insulator and itinerant electron limit and it is a cubic spinel, where the V atoms form a pyrochlore lattice of corner-sharing tetrahedra and as a matter of fact, antiferromagnetic (AF) interactions between the V atoms are highly frustrated. At $T_S = 51$ K, $ZnV_2O_4$ undergoes a structural phase transition from cubic to tetragonal (c/a < 1) with a compression of the $VO_6$ octahedron along the $c$ axis [1] and the system possibly orbitally orders. This lifts the geometrical frustration of the cubic phase and a second transition at $T_N = 40$ K occurs which is of a magnetic nature and the system orders antiferromagnetically [1-3]. The lattice formed by the V atoms can be described as built up by V-V chains running along the [110], [011], and [101] directions. The magnetic structure, found by neutron diffraction [1,4] is AF along the [110] direction (within the $ab$ plane), but along the [101] and [011] (off plane) directions the V moments order. Moreover, the effective hopping and the $d$-$d$ transfer integral of $ZnV_2O_4$ estimated from the X-ray photoemission spectra are close to the metallic $LiV_2O_4$ [5].

Furthermore, paramagnetic metals and Mott-Hubbard insulators represent two fundamentally different phases that can be interchanged by increasing or decreasing electronic correlations through a first-order quantum phase transition (QPT).[6] Nevertheless, it is highly challenging to characterize the electronic properties of materials when approaching the quantum phase transition (QPT) either from Mott insulator side or from the itinerant electron side. Among very few materials, $AV_2O_4$ spinels are a family of Mott insulators that fulfil the criterion because of the metal-metal separation can be changed by applying chemical pressure i.e., by changing the size of the $A^{2+}$ cation [3]. The absence of $e_g$ electrons makes direct V-V hybridization between

$t_{2g}$ orbitals the only relevant contribution to the hopping amplitude. Therefore, by proper doping on A-site $ZnV_2O_4$ may be transferred to the Mott transition.

Moreover, it has been shown that in $ZnV_2O_4$ away from the strong-coupling regime *bond and magnetic* ordering are independent of each other [7]. If the bond ordering is strengthened away from the strong-coupling regime then it may correspond to ferroelectricity as is observed in $CdV_2O_4$ [8]. It has also been suggested by Kuntscher et al. [7] that ferroelectricity can be increased significantly by decreasing V-V distance as the structural distortion observed in $ZnV_2O_4$ is identical to $CdV_2O_4$. Furthermore, the main limiting factors for $ZnV_2O_4$ to show spontaneous polarization are the losses or leak currents. The increase in the value of charge gap reduces the losses. In $ZnV_2O_4$ when external pressure is applied, which lead to the decrease in V-V distance, the charge gap increases above a critical pressure [7]. As the V-V separation decreases the system moves towards the itinerant electron limit [3]. Therefore, closing the Mott gap of this geometrically frustrated Mott insulator is a potential route for enhancing ferroelectricity which could be achieved in $ZnV_2O_4$ by reducing the V-V distance. Chemical substitution is the best way in doing this.

Moreover, when Co is doped in $ZnV_2O_4$ the system moves towards the itinerant electron limit [9]. In the present paper we have varied the chemical pressure in $ZnV_2O_4$ by doping Co on the Zn site and studied the effect of it on local structure surrounding V sites by carrying out Extended X-ray analysis of fine structure (EXAFS) measurements at V K-edge. The EXAFS results establish the mechanism of Co substitution in the $ZnV_2O_4$ lattice in a microscopic scale.

**EXPERIMENTAL**:

Polycrystalline ($Zn_{0.9}Co_{0.1}V_2O_4$) were prepared by solid state reaction route. ZnO, CoO and $V_2O_3$ were mixed in appropriate ratios and pressed into pellets. The pellets were sealed in quartz tube under high vacuum and heat treated in a furnace at 800°C for 60 h. X-ray diffraction measurements were performed at 10 K and at room temperature. Magnetic measurements were carried out with Vibrating Sample Magnetometer (VSM). Synchrotron XRD measurements were performed at angle dispersive X-ray diffraction (ADXRD) beamline (BL-12) [10] on Indus-2 synchrotron source [11]. The beamline consists of a Si (111) based double crystal monochromator and two experimental stations namely a six circle diffractometer with a scintillation point detector and Mar 345 dtb image plate area detector. SXRD measurements were carried out using the image plate. The X-ray wave length used for the present study was accurately calibrated by doing X-ray diffraction on LaB6 NIST standard. Data reductions were done using Fit2D software [12]. EXAFS measurements of these samples at V K-edge have been carried out in transmission mode at the Scanning EXAFS Beamline (BL-9) at the INDUS-2 Synchrotron Source (2.5 GeV, 100 mA) at the Raja Ramanna Centre for Advanced Technology (RRCAT), Indore, India [13]. The beamline uses a double crystal monochromator (DCM) which works in the photon energy range of 4-25 KeV with a resolution of $10^4$ at 10 KeV. A 1.5 m horizontal pre-mirror with meridonial cylindrical curvature is used prior to the DCM for collimation of the beam and higher harmonic rejection. The second crystal of DCM is a sagittal cylinder used for horizontal focussing while a Rh/Pt coated bendable post mirror facing down is used for vertical focusing of the beam at the sample position. For the present set of samples EXAFS measurements at V K-edge has been carried out in transmission mode where, the sample is placed between two ionization chamber detectors.

The first ionization chamber measures the incident flux ($I_0$) and the second ionization chamber measures the transmitted intensity ($I_t$) and the absorbance of the sample is obtained as ($\mu = \exp(-\frac{I_t}{I_0})$). The EXAFS spectra of the sample at V K-edge were recorded in the energy range (5400-6112) eV.

**RESULTS & DISCUSSIONS**:

Synchroton X-ray diffraction was carried out on polycrystalline $Zn_{0.9}Co_{0.1}V_2O_4$ at different temperatures to refine the crystal structure and thereby obtain structural parameters. Figure 1 shows the representative diffraction pattern and the Rietveld fit. All the samples were found to be of single phase, iso-structural and crystallize in space group *Fd-3m*. In this space group, the $V^{3+}$ site is octahedrally coordinated with oxygen, leading to low lying partially empty $t_{2g}$ levels. This leads to orbital degrees of freedom in addition to the $S = 1$ moment on the $V^{3+}$ ion. The *A* ions are tetrahedrally coordinated with oxygen, leading to low lying $e_g$ orbitals. The variations of cell parameters, V-V distances and V-O bond lengths with temperature are shown in Fig. 2. All the three parameters show non-monotonic temperature dependence and a minimum at ~40K is observed.

Figure 3 shows the temperature variation of magnetization of the $Zn_{0.9}Co_{0.1}V_2O_4$ samples under field-cooled (FC) and zero-field-cooled (ZFC) conditions. It can be seen from the above figure that the sample undergoes an anti-ferromagnetic transition at 40 K whereas for pure $ZnV_2O_4$ the transition temperature is ~42K [9]. The field dependence of magnetization, M(H) as shown in the inset of Fig.3 indicates the anti-ferromagnetic ordering in Co doped $ZnV_2O_4$. Moreover, in the M(T) curve of $ZnV_2O_4$ a transition at ~40K is observed [1,9], which is due to the structural transition. But in the Co doped sample, the second transition does not appear. The bifurcation of the ZFC and FC magnetization curves observed in these compounds possibly arises from the frustrated behavior in the spinel vanadates.

In order to investigate the local structure around V atoms in Co-doped $ZnV_2O_4$ at room temperature and at 10K we have made the Synchrotron based Extended X-ray Absorption Fine Structure (EXAFS) study. Figure 4 represents the experimental EXAFS ($\mu(E)$ versus $E$) spectra of undoped and Co-doped $ZnV_2O_4$ respectively measured at V K-edge. In order to take care of the oscillations in the absorption spectra, the energy dependent absorption coefficient $\mu(E)$ has been converted to absorption function $\chi(E)$ defined as follows [14]:

$$\chi(E) = \frac{\mu(E) - \mu_0(E)}{\Delta\mu_0(E_0)} \qquad (1)$$

where $E_0$ is the absorption edge energy, $\mu_0(E_0)$ is the bare atom background and $\Delta\mu_0(E_0)$ is the step in the $\mu(E)$ value at the absorption edge. After converting the energy scale to the photoelectron wave number scale ($k$) as defined by,

$$k = \sqrt{\frac{2m(E-E_0)}{\hbar^2}} \qquad (2)$$

the energy dependent absorption coefficient $\chi(E)$ has been converted to the wave number dependent absorption coefficient $\chi(k)$, where $m$ is the electron mass. Finally, $\chi(k)$ is weighted by $k$ to amplify the oscillation at high $k$ and the $\chi(k)k$ functions are Fourier transformed in $R$ space to generate the $\chi(R)$ versus $R$ (or FT-EXAFS) spectra in terms of the real distances from the center of the absorbing atom. it should be mentioned here that a set of EXAFS data analysis program available within the IFEFFIT software package have been used for reduction and fitting of the experimental EXAFS data [15]. This includes data reduction and Fourier transform to derive the $\chi(R)$ versus $R$ spectra from the absorption spectra (using ATHENA software), generation of the theoretical EXAFS spectra starting from an assumed crystallographic structure and finally fitting of the experimental data with the theoretical spectra using the FEFF 6.0 code (using ARTEMIS software). The goodness of the fit in the above process is generally expressed by the $R_{factor}$ factor which is defined as [16]:

$$R_{factor} = \sum \frac{[Im(\chi_{dat}(r_i) - \chi_{th}(r_i))]^2 + [Re(\chi_{dat}(r_i) - \chi_{th}(r_i))]^2}{[Im(\chi_{dat}(r_i))]^2 + [Re(\chi_{dat}(r_i))]^2} \qquad (3)$$

where, $\chi_{dat}$ and $\chi_{th}$ refer to the experimental and theoretical $\chi(r)$ values respectively and Im and Re refer to the imaginary and real parts of the respective quantities.

Fig. 5 represents $\chi(R)$ versus $R$ spectra of undoped and doped $ZnV_2O_4$ samples respectively at V K-edge along with the best fit theoretical spectra. For undoped and Co-doped $ZnV_2O_4$ samples the fittings have been carried out by assuming cubic $ZnV_2O_4$ structure. It should be noted here that as found by M. Reehuis et. al. [17], $ZnV_2O_4$ shows cubic to tetragonal structural phase transition at 40K. We have also seen the structural transition from the structural and magnetic properties study [9]. But as we have discussed above no structural transition is observed when Co is doped in $ZnV_2O_4$. As a matter of fact, for comparison both undoped and doped $ZnV_2O_4$ the EXAFS experimental spectra fitted with the cubic structure and has resulted in good fit with low R-factor. The different paths required for theoretical fitting of the EXAFS data have been obtained from the crystal structure obtained for cubic doped and undoped $ZnV_2O_4$ from XRD measurements.

For undoped and Co doped $ZnV_2O_4$ samples (Fig. 5), the first and second major peaks in the radial distribution function correspond to the nearest oxygen and the V/Zn atoms respectively from the central V atom. The theoretical FT-EXAFS spectra have been generated assuming the first oxygen shell (V-O) at 1.98Å with coordination number (CN) of 6, second V shell (V-V) at 2.9Å having CN of 6, third Zn shell (V-Zn) at 3.32Å with CN of 6 and fourth oxygen shell (V-O) at 3.81Å .The data have been fitted in the k range of (3-10.5Å$^{-1}$) and upto (3.5 Å) in R space.

The best fit parameters of the above fittings have been shown in Table-1 from where it can be observed that the oxygen coordination of the first V-O shell does not change upon Co doping, which shows that Co is going into the Zn sites of the lattice as $Co^{2+}$ without causing any charge imbalance. It can also be seen from Table-1 that both for first and fourth shell there is no change in the V-O distances with Co doping. This may happen only if there is no difference between the ionic radii of $Zn^{2+}$ and $Co^{2+}$ and this is the case when $Co^{2+}$ exists in the lattice in high spin state. At high spin state the ionic radius of $Co^{2+}$ is 88.5 pm whereas at low spin state it is 74 pm. On the other hand, the ionic radius of $Zn^{2+}$ is 88 pm. Therefore, it is observed from the EXAFS measurement that Co ion is in the high spin state which is also supported by the increased magnetization in the samples with Co doping. This also causes very small changes in V-V and V-Zn distances with Co-doping. Small decrease in V-V distance and increase in V-Zn distance are observed which is due to the slight higher value in ionic radius of Co ion.

It is also observed from Fig. 2 that there is an obvious effect around magnetic transition in lattice parameter and bond lengths. From Table 1 also it is found that there is a change in the bond lengths in 2$^{nd}$ and 3$^{rd}$ and 4$^{th}$ shell. It is found that in the 2$^{nd}$ shell there is a change in V-V length from 2.89Å (at 300 K) to 2.84 Å (at 10K), in the 3$^{rd}$ shell a change in V-Zn distance from 3.32Å (at 300K) to 3.36Å (at 10K) and in the 4$^{th}$ shell a change in V-O length from 3.8Å (at

300K) to 3.75Å (at 10K) are observed. Thus, these results clearly indicate a change in local structural parameters around V atoms across the magnetic transition temperature in $ZnV_2O_4$. The EXAFS results suggest that increase in magnetization with Co doping is due to the existence of Co is in high spin state.

**CONCLUSION**:

Co doped $ZnV_2O_4$ has been investigated by Synchrotron X-ray diffraction, Magnetization measurement and Extended X-ray absorption fine structure measurements. With Co doping in the Zn site the system moves towards the itinerant electron limit. Moreover, with Co doping the magnetization value also is increased. V-O and V-V bond lengths and the lattice parameters obtained from the Synchrotron X-ray diffraction show non-monotonic behavior around the magnetic transition temperature. No structural transition is observed when Co is doped in $ZnV_2O_4$. From the EXAFS study it is found that there is no change in the oxygen coordination in first V-O shell which proves that Co goes into $Zn^{2+}$ sites as $Co^{2+}$. It is also observed that V-O bond length does not change with Co doping which indicates that Co exists in the High spin state in $ZnV_2O_4$ lattice. However, bond lengths around V sites do change below the magnetic transition temperature which indicates a change in local structural parameters around V atoms across the magnetic transition temperature in Co doped $ZnV_2O_4$.

**ACKNOWLEDGEMENT**

SC is grateful to BRNS, DAE ((Grant No.: 2013/37P/43/BRNS), DST (Grant No.: SR/S2/CMP-26/2008) and CSIR (Grant No.: 03(1142)/09/EMR-II) for providing financial support. PS is grateful to CSIR for providing financial support.

**Table-1: Parameters obtained from EXAFS measurement:**

|  |  | $ZnV_2O_4$ | $ZnV_2O_4$ (10K) | $Zn_{0.9}Co_{0.1}V_2O_4$ | $Zn_{0.9}Co_{0.1}V_2O_4$ (10K) |
|---|---|---|---|---|---|
| V-O(Å) | CN(6) | 4.35±0.45 | 4.02±0.36 | 4.32±0.36 | 4.34±0.28 |
|  | R | 1.98±0.009 | 1.93±0.006 | 1.98±0.007 | 1.99±0.006 |
|  | $\sigma^2$ | 0.003±0.001 | 0.003±0.0009 | 0.003±0.001 | 0.003±0.0009 |
| V-V(Å) | CN(6) | 4.91±0.84 | 4.9±0.45 | 6 | 6±0.49 |
|  | R | 2.9±0.018 | 2.9±0.007 | 2.89±0.01 | 2.84±0.02 |
|  | $\sigma^2$ | 0.01±0.003 | 0.008±0.001 | 0.008±0.001 | 0.008±0.0008 |
| V-Zn(Å) | CN(6) | 7.2±1.77 | 7.2±1.74 | 6 | 5.76±1.5 |
|  | R | 3.32±0.03 | 3.33±0.026 | 3.32±0.043 | 3.36±0.019 |
|  | $\sigma^2$ | 0.017±0.004 | 0.017±0.003 | 0.014±0.008 | 0.014±0.002 |
| V-O(Å) | CN(6) | 6±1.25 | 6±1.25 | 8.34±2.1 | 7.9±2.433 |

|   | R | 3.81±0.028 | 3.76±0.011 | 3.8±0.024 | 3.75±0.019 |
|---|---|---|---|---|---|
|   | $\sigma^2$ | 0.001±0.007 | 0.001±0.001 | 0.002±0.006 | 0.005±0.004 |
|   | $R_{factor}$ | 0.018 | 0.013 | 0.02 | 0.008 |

**Figure Captions:**

Figure 1. Synhrotron X-ray diffraction pattern with Reitveld refinement for $Zn_{0.9}Co_{0.1}V_2O_4$ sample at temperature range 18K - 300K.

Figure 2. Change of a(Å), V-V(Å) and V-O(Å) parameters with respect to temperature.

Figure 3. Temperature variation of magnetization for $Zn_{0.9}Co_0V_2O_4$ spinel at H=500 Oe. Inset: M- Curve at 10 K for $Zn_{0.9}Co_0V_2O_4$.

Figure 4. Normalized experimental V K-edge EXAFS ($\mu(E)$ versus E) spectra for undoped and doped $ZnV_2O_4$ samples measured at room temperature and at 10K.

Figure 5. The experimental $\chi(R)$ versus $R$ plots and the theoretical fits of undoped and Co doped $ZnV_2O_4$ samples at V K-edge.

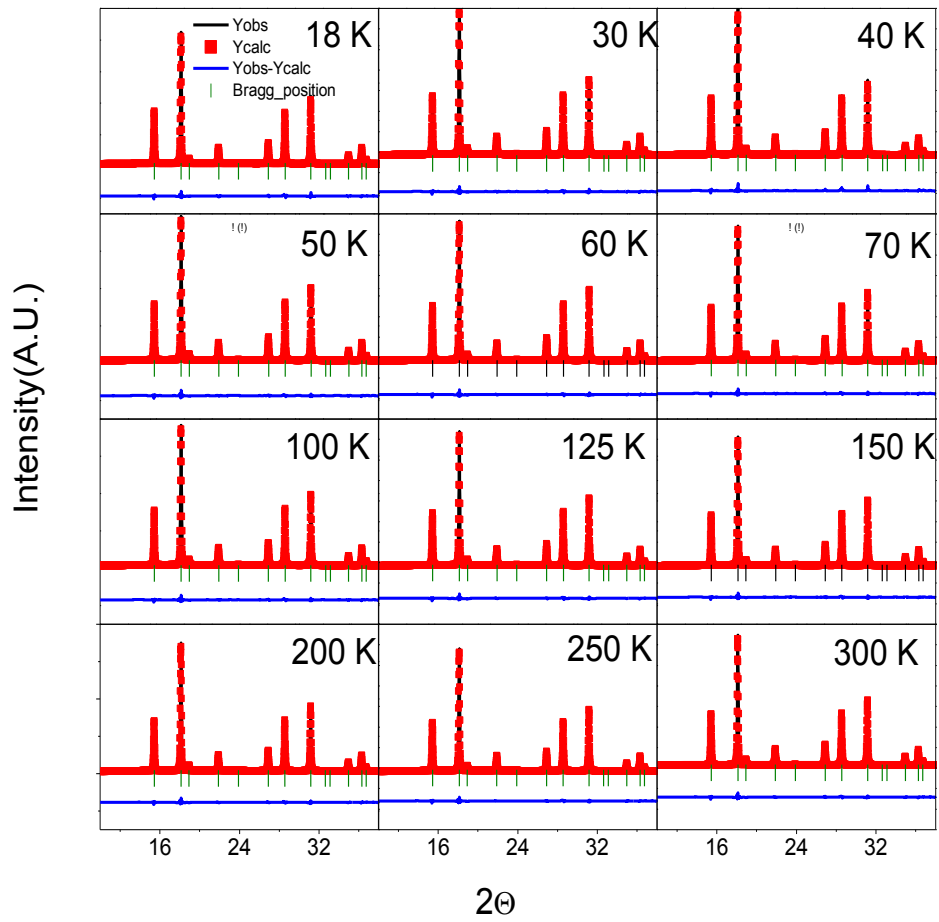

Figure 1

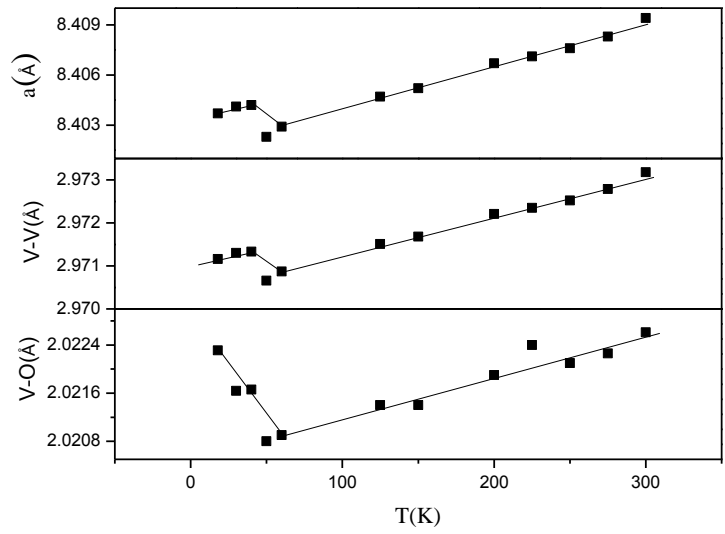

Figure 2

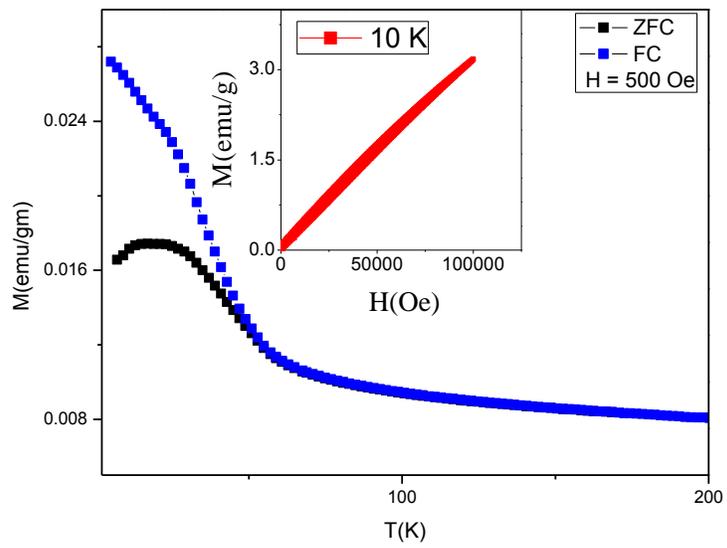

Figure 3

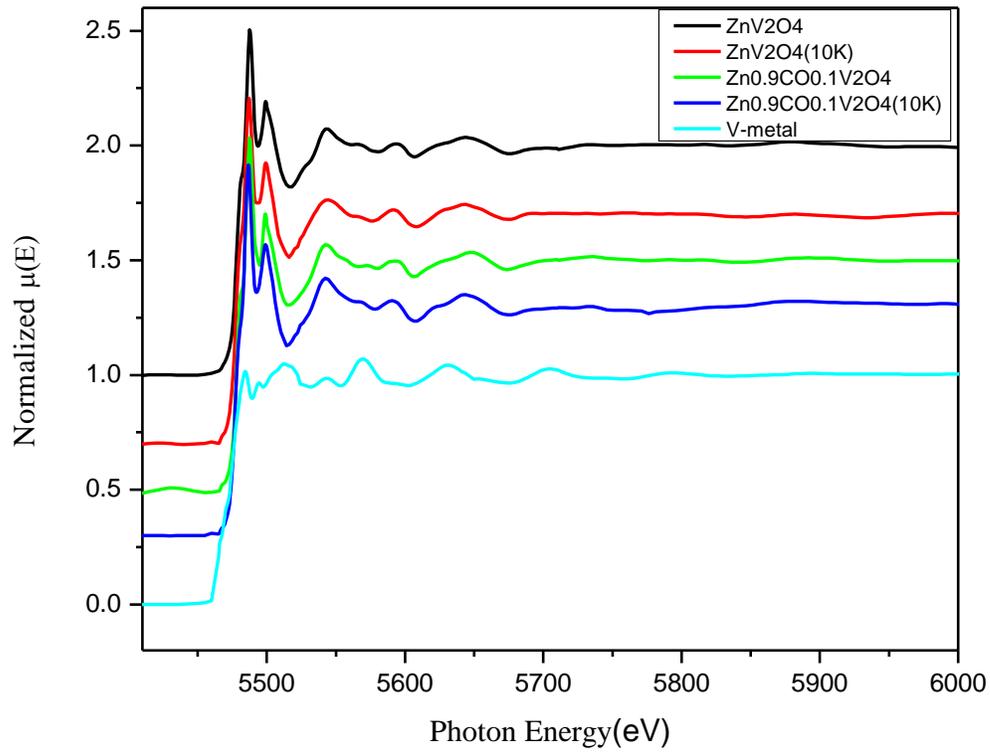

Figure 4

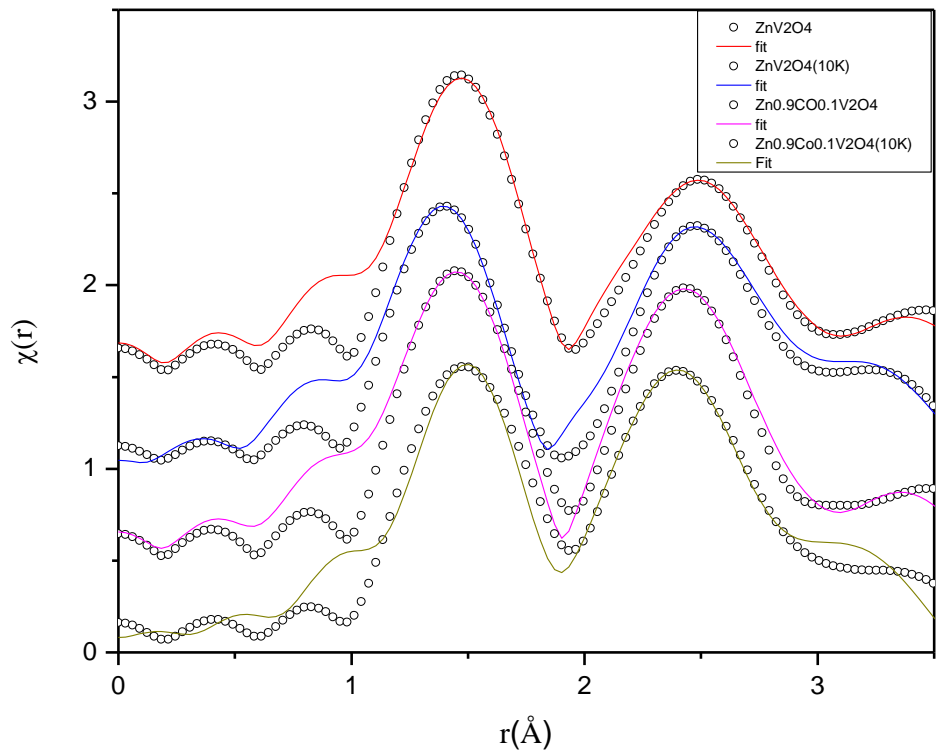

Figure 5